\documentclass[11pt,a4paper]{article}
\usepackage{jheppub}

\usepackage[utf8]{inputenc}

\usepackage{mciteplus}
\usepackage{hyperref}
\usepackage{amssymb}
\usepackage{amsthm}
\usepackage{amstext}
\usepackage{amsmath}
\usepackage{amsfonts}
\usepackage{dsfont}
\usepackage{cancel}
\usepackage{wrapfig}
\usepackage{graphicx}
\usepackage{color}

\usepackage{slashed}

\usepackage{enumitem}

\usepackage{xparse}

\newcommand \vev [1] {\langle{#1}\rangle}

\def\eps{\epsilon}

\allowdisplaybreaks[3]

\makeatletter
\renewcommand\@fpheader{} 
\renewcommand\@journal{}
\makeatother

\title{The two-loop five-particle amplitude in $\mathcal{N}=8$ supergravity}

\author[a]{Dmitry~Chicherin}
\author[b]{Thomas~Gehrmann}
\author[a]{Johannes~M.~Henn}
\author[c]{Pascal~Wasser}
\author[a]{Yang~Zhang}
\author[a]{Simone~Zoia}
\affiliation[a]{Max-Planck-Institut für Physik,
  Werner-Heisenberg-Institut, D-80805 M\"unchen, Germany}
\affiliation[b]{Physik-Institut, Universität Z\"urich, Wintherturerstrasse 190, CH-8057 Z\"urich, Switzerland}
\affiliation[c]{PRISMA Cluster of Excellence, Johannes Gutenberg University, D-55128 Mainz, Germany}

\emailAdd{chicheri@mpp.mpg.de}
\emailAdd{thomas.gehrmann@uzh.ch}
\emailAdd{henn@mpp.mpg.de}
\emailAdd{wasserp@uni-mainz.de}
\emailAdd{yzhang@mpp.mpg.de}
\emailAdd{zoia@mpp.mpg.de}

\abstract{We compute for the first time the two-loop five-particle amplitude in \mbox{$\mathcal{N}=8$} supergravity. 
Starting from the known integrand, we perform an integration-by-parts reduction and express the answer
in terms of uniform weight master integrals. The latter are known to evaluate to non-planar pentagon
functions, described by a 31-letter symbol alphabet.  We express the final result for the amplitude in terms of 
uniform weight four symbols, multiplied by a small set of rational factors. 
The amplitude satisfies the expected factorization properties when one external graviton becomes soft, and when two 
external gravitons become collinear.
We verify that the soft divergences of the amplitude exponentiate, and extract the finite remainder function.
The latter depends on fewer rational factors, and is independent of one of the symbol letters.
By analyzing identities involving rational factors and symbols we find a remarkably compact representation in terms of
a single seed function, summed over all permutations of external particles.
Finally, we work out the multi-Regge limit, and present explicitly the leading logarithmic terms in the limit.
The full symbol of the IR-subtracted hard function is provided as an ancillary file.
}


\begin{document}

\begin{flushright}\begin{tabular}{r}
\end{tabular}\end{flushright}
\vspace{-14.1mm}
\maketitle

\clearpage

\section{Introduction}

The last decades have seen remarkable progress in our understanding of 
scattering amplitudes in gauge and gravity theories. Among the different theories,
the ones with maximal degree of supersymmetry, $\mathcal{N}=4$ super Yang-Mills (sYM),
and $\mathcal{N}=8$ supergravity, are expected to be the simplest \cite{ArkaniHamed:2008gz}.
They have proven to be a fantastic laboratory to
explore properties of quantum field theory. Studies in these theories have stipulated
advances in our understanding of infrared divergences, Regge limits, symmetry properties,
dualities, connections to string theory, loop integrands, 
special functions arising from Feynman integrals, symbols,
and many other properties of scattering amplitudes.

Many studies in these theories dealt with properties of tree-level amplitudes 
and loop integrands. This is particularly interesting, as the latter encodes,
sometimes in a very concrete way, properties of the answer after integration.
For example, representations of loop integrands having manifest ultraviolet (UV)
properties may help answer the question whether $\mathcal{N}=8$ supergravity is
perturbatively UV finite~\cite{Bern:2017yxu,Bern:2017ucb}. The analysis of leading singularities \cite{Cachazo:2008vp,Cachazo:2008dx,ArkaniHamed:2010gh}, i.e. maximal residues
of integrands, is closely linked to the rational functions appearing after integration.
Moreover, there is a conjectured relation between Feynman integrals having so-called
{\it dlog} integrands, and iterated integrals of uniform weight \cite{ArkaniHamed:2010gh,Henn:2013pwa,Arkani-Hamed:2014via}.
In all these studies, having perturbative `data', i.e. explicit results for scattering amplitudes,
was invaluable. In ${\mathcal{N}}=4$ super Yang-Mills, a wealth of perturbative data is
available. However, such data is particularly sparse for ${\mathcal N}=8$ super Yang-Mills at
the integrated level. Up to now, beyond one-loop, only the two-loop four-particle amplitude
is known~\cite{Naculich:2008ew,Brandhuber:2008tf,BoucherVeronneau:2011qv}. 

Very recently, conceptual and technical progress in integration-by-parts relations \cite{vonManteuffel:2014ixa,Peraro:2016wsq,Bern:2017gdk,Kosower:2018obg,Maierhofer:2018gpa,Boehm:2018fpv,Chawdhry:2018awn} and in
evaluating Feynman integrals via differential equations~\cite{Gehrmann:1999as,Henn:2013pwa} culminated in
the evaluation of all planar~\cite{Gehrmann:2015bfy,Papadopoulos:2015jft,Gehrmann:2018yef} and non-planar~\cite{Abreu:2018rcw,Chicherin:2018mue,Abreu:2018aqd,Chicherin:2018old,Chicherin:2018yne} Feynman integrals required for two-loop five-particle scattering amplitudes.
The corresponding functions, dubbed pentagon functions, fall into a class of iterated integrals that are described by an alphabet of $31$ logarithmic integration kernels called letters \cite{Chicherin:2017dob}. This alphabet is closely linked to the (actual and spurious) singularities of
the pentagon functions.
At the planar level, all two-loop Yang-Mills scattering amplitudes have been evaluated,
numerically~\cite{Badger:2017jhb,Abreu:2018jgq} and analytically~\cite{Gehrmann:2015bfy,Dunbar:2016aux,Badger:2018enw,Abreu:2018zmy}.
The very recent results on the non-planar integrals allowed the analytic evaluation of the symbol of the full-color two-loop five-point $\mathcal{N}=4$ sYM amplitude~\cite{Abreu:2018aqd,Chicherin:2018yne}.
In this paper, we supply more such data, by computing the symbol of the two-loop five-graviton amplitude in ${\mathcal{N}}=8$ supergravity. 

The comparison between the sYM and supergravity theories is a very interesting one, as they have
many similarities, but also important differences. For example, while in sYM the concept of color-ordering
and `t Hooft expansion is fundamental, the same does not exist in supergravity amplitudes.
This implies that the latter are intrinsically non-planar. Moreover, the supergravity
amplitudes have a permutation symmetry under exchange of any of the external gravitons.
This property is typically not obvious for individual Feynman diagrams or intermediate expressions, and
as a consequence certain simple properties of the final answer sometimes appear only after adding up
all contributions to an amplitude.

A good example of this fact are the infrared properties of (super)gravity. 
It is well-known that perturbative gravity has a simpler infrared structure as compared to Yang-Mills theories. 
Its scattering amplitudes are in fact free of collinear divergences~\cite{Weinberg:1965nx}.
An intuitive explanation is given by the fact that, already at classical level, gravitational radiation in the forward direction of the emitter is suppressed with respect to, for example, electro-magnetic radiation~\cite{VanNieuwenhuizen:1973qf}.
The absence of collinear divergences can be proven more formally using power counting arguments~\cite{Akhoury:2011kq}, or within the SCET formalism~\cite{Beneke:2012xa}.

On the other hand, just like Yang-Mills amplitudes, graviton amplitudes have soft singularities, so that one expects them to have a single pole in the dimensional regulator $\eps$ per loop order (where $D=4-2 \eps$). In particular, it was found in refs.~\cite{Weinberg:1965nx, Dunbar:1995ed, Naculich:2008ew, Naculich:2011ry, White:2011yy, Akhoury:2011kq,Beneke:2012xa} that the soft divergences exponentiate in a remarkably simple way. 
The infrared structure of graviton amplitudes is therefore much simpler compared to non-Abelian gauge theories~\cite{Almelid:2015jia} or even QED.

In recent years, the structure of scattering amplitudes at subleading orders in the soft limits has received a lot of attention. 
While the subleading soft theorem is expected to be exact at tree-level and at the level of (four-dimensional) loop integrands, 
at the integrated level there may be specific correction terms \cite{Cachazo:2014fwa,Bern:2014oka,Bern:2014vva}.
In particular, there is an anomalous term at one loop. Due to the fact that the coupling is dimensionful, the latter is expected to be one-loop exact. It would be interesting to test this prediction. 

The outline of the paper is as follows. In section \ref{sec:kinematics}, we review the kinematics of five-particle scattering 
and of the relevant non-planar pentagon function space. Section \ref{sec:examples} is dedicated to reviewing
the previously known tree-level and one-loop amplitudes, while we discuss the structure of 
infrared divergences in section \ref{sec:divergences}. Section \ref{sec:calculation} explains our calculation of the two-loop amplitude,
with the main result given in section \ref{sec:mainresult}. In section \ref{sec:limits} we analyze the soft and collinear as well as the multi-Regge limit
of our result. We draw our conclusions in section \ref{sec:conclusion}.


\section{Kinematics and pentagon functions}
\label{sec:kinematics}

The scattering of five massless particles carrying momenta $p_i^{\mu}$ is described by five independent Mandelstam invariants, $s_{12}$, $s_{23}$, $s_{34}$, $s_{45}$, $s_{51}$, with $s_{ij} = 2 p_i \cdot p_j$, and the pseudo-scalar $\epsilon_5 = \text{tr} [\gamma_5 \slashed{p}_4 \slashed{p}_5 \slashed{p}_1 \slashed{p}_2 ]$. The square of the latter is a scalar, and can therefore be expressed in terms of the $s_{ij}$. This can be done through $\Delta = (\epsilon_5)^2$, where $\Delta$ is the Gram determinant $\Delta = |2 p_i \cdot p_j|$, with $1\le i,j \le 4$.

The Feynman integrals relevant for the scattering of five massless particles up to two loops evaluate to a special class of polylogarithmic functions called pentagon functions \cite{Chicherin:2017dob}. They can be expressed as iterated integrals of the form $\int d\log W_{i_1}...\int d\log W_{i_n}$, where the $W_i$ are algebraic functions of the kinematics called letters, and the number of integrations $n$ defines the transcendental weight of the function. The letters encode the branch-cut structure of the integrals, and their ensemble $\{ W_i \}$ is called alphabet.

The $\mathbb{Q}$-linear combinations of $d\log$ iterated integrals of the same transcendental weight are called pure functions of uniform weight. They are the natural ingredients in the analytic expressions of scattering amplitudes. Instead of working directly with the iterated integrals, we will consider their symbols. The \textit{symbol} $\cal S$~\cite{Goncharov:2010jf,Duhr:2011zq} maps a $d\log$ iterated integral into a formal sum of the ordered sets of its $d\log$ kernels 
\begin{align}
\sum_{i_1,\ldots,i_n} c_{i_1,\ldots,i_n} \int d\log W_{i_1}\ldots\int d\log W_{i_n}  \;\;\overset{\cal S}{\longrightarrow}\;\; \sum_{i_1,\ldots,i_n} c_{i_1,\ldots,i_n} [W_{i_1},\ldots,W_{i_n}] \, ,
\end{align}
where $c_{i_1,\ldots,i_n}$ are rational constants.
The symbols capture all the combinatorial and analytic properties of the corresponding functions, but they are not sufficient for the numerical evaluation of the integrals, since information about the integration contours is omitted. 

The 31 letters $\{W_i\}_{i=1}^{31}$ of the pentagon alphabet, defined in eqs.~(2.5) and~(2.6) of \cite{Chicherin:2017dob}, have well defined transformation rules under parity conjugation.  There are 26 parity-even and 5 parity-odd letters
\begin{align}
& d \log (W_i)^* = + d \log W_i \, , \qquad i=1,\ldots,25 , 31 \, , \notag \\
& d \log (W_i)^* = - d \log W_i \, ,  \qquad i=26,\ldots,30 \, . 
\end{align}

The first entry of the symbol encodes the discontinuities of the corresponding function. The symbols entering scattering amplitudes are therefore subject to first entry conditions due to physical constraints on the allowed discontinuities, which can occur only where two-particle Mandelstam invariants $s_{ij}$ vanish. For the pentagon alphabet, we have that  $\{ s_{ij} \}_{1 \leq i<j \leq 5} = \{ W_i\}_{i=1}^5 \cup \{ W_i \}_{i=16}^{20}$, from which it follows that only the latter subset of 10 letters is allowed in the first entries.

Of the remaining letters, $\{W_{i}\}_{i=6}^{15} \cup \{W_{i}\}_{i=21}^{25}$ are given by simple linear combinations of $s_{ij}$, which can be obtained from cyclic permutations of $s_{13}$; the five parity-odd letters $\{W_{i}\}_{i=26}^{30}$ are pure phases, and they mix with the parity-even ones under permutations; finally, the last letter is the pseudo-scalar $W_{31} = \epsilon_5$, with $d\log W_{31}$ invariant under permutations.
 
In addition to the first entry condition, it was first conjectured~\cite{Chicherin:2017dob} and then confirmed~\cite{Abreu:2018aqd,Chicherin:2018old} that certain pairs of letters do not appear as first and second entries of the symbols. This constraint, referred to as second entry condition, is an observation. It would be interesting to find its physical motivation.

\section{Five-graviton scattering amplitudes: tree-level and one-loop cases}
\label{sec:examples}

It is instructive to start by reviewing the known lower-order results. 
We may hope to infer from these expressions some educated guesses about the structure of the amplitudes at higher loop orders.

We expand the amplitudes in the gravitational coupling constant $\kappa$, with $\kappa^2 = 32 \pi G$,
\begin{align}
M_{5} = \delta^{(16)}(Q)\,   \sum_{\ell \ge0 }  \bigg(\frac{\kappa}{2}\bigg)^{2 \ell + 3} \, \left( \frac{e^{- \eps \gamma_{\rm E} }}{(4 \pi)^{2-\eps}} \right)^{\ell} \, \mathcal{M}^{(\ell)}_{5} \,,
\end{align}
where $\delta^{(16)}(Q)$ is the super-momentum conservation delta function. 
Note that $\kappa$ has dimension of $1/p$.
An expression of the tree-level amplitude following from the Kawai-Lewellen-Tye relations~\cite{Kawai:1985xq} is given by~\cite{Berends:1988zp}
\begin{align}
\label{eq:M_tree}
\mathcal{M}^{(0)}_{5} = - s_{12} s_{34} \text{PT}(12345) \text{PT}(21435) -
	  s_{13} s_{24} \text{PT}(13245) \text{PT}(31425) \, ,
\end{align}
where we introduced the Parke-Taylor (PT) tree-level factor 
\begin{align}
\text{PT}(i_1 i_2 i_3 i_4 i_5) = \frac{1}{\vev{i_1 i_2}\vev{i_2 i_3}\vev{i_3 i_4}\vev{i_4 i_5}\vev{i_5 i_1}} \, .
\end{align}
We note that, although not obvious, the expression in eq.~\eqref{eq:M_tree} is fully symmetric under permutation of the external legs.
This is related to the following property. The rational factors appearing in (\ref{eq:M_tree}) are of the form
\begin{align}
\label{eq:treePTPT}
s_{ij} s_{kl} \text{PT}(\sigma) \text{PT}(\rho) \, ,
\end{align}
where Greek letters $\sigma$ and $\rho$ denote arbitrary permutations of $(12345)$. 
These factors satisfy many relations, and only 146 of them are linearly independent. 
It is precisely these relations that allow for the permutation symmetry.  
As a result, we may write the tree-level amplitude equivalently in the manifestly symmetric form
\begin{align}
\label{eq:M_tree2}
\mathcal{M}^{(0)}_{5} = \frac{1}{60} \, \sum_{S_5} \bigl[ - s_{12} s_{34} \text{PT}(12345) \text{PT}(21435) \bigr]  \, ,
\end{align}
where the sum runs over the $5!=120$ permutations of the external legs.

A new class of rational factors appears in the one-loop amplitude, which can be written as \cite{Bern:1998sv}
\begin{align}
\label{eq:M_1loop}
\mathcal{M}^{(1)}_{5} =  - \sum_{S_5} \Bigl[ & s_{45} s_{12}^2 s_{23}^2 \text{PT}(12345) \text{PT}(12354) \mathcal{I}_4^{(45)} 
  + 2 \epsilon \frac{[12][23][34][45][51]}{\vev{1 2}\vev{2 3}\vev{3 4}\vev{4 5}\vev{5 1}} \mathcal{I}_5^{6-2 \epsilon} \Bigr]\, ,
\end{align}
where $\mathcal{I}_4^{(45)}$ is the scalar one-mass box with external momenta $p_1,p_2,p_3$ and $p_4+p_5$, and $\mathcal{I}_5^{6-2 \epsilon}$ is the massless scalar pentagon in $D=6-2\epsilon$ dimensions.
For the present discussion, we will not need the explicit expressions of $\mathcal{I}_4^{(45)}$ and $\mathcal{I}_5^{6-2 \epsilon}$, but
will just note that they evaluate to pure functions, with overall rational prefactors $1/(s_{12} s_{23})$ and $1/\epsilon_5$, respectively.  Taking this information into account, we see from eq.~\eqref{eq:M_1loop} that the integrated amplitude will depend on two classes of rational factors. 
On the one hand, there is the one-loop generalization of the tree-level factors~\eqref{eq:treePTPT}, namely certain permutations of
\begin{align}
\label{eq:1loopPTPT}
s_{ij} s_{kl} s_{mn} \text{PT}(\sigma) \text{PT}(\rho) \,.
\end{align}
The latter form a 290-dimensional space over $\mathbb{Q}$. 
On the other hand, the six-dimensional pentagon integral introduces a new object,
\begin{align}
\label{eq:extraPrefactor1loop}
\frac{1}{\epsilon_5}\frac{[12][23][34][45][51]}{\vev{1 2}\vev{2 3}\vev{3 4}\vev{4 5}\vev{5 1}}\, ,
\end{align}
which is linearly independent of the factors in~(\ref{eq:1loopPTPT}) and, quite remarkably, is permutation invariant. It is interesting that this factor enters the amplitude only at $\mathcal{O}(\eps)$.

The one-loop amplitude offers other good examples of non-trivial relations existing among these rational functions. 
For example, the prefactor coming from the one-mass box in eq.~\eqref{eq:M_1loop} vanishes upon summing over all its $S_5$ permutations
\begin{align}
\sum_{S_5} s_{12} s_{23} s_{45} \, \text{PT}(12345) \text{PT}(12354) = 0 \,.
\end{align}
Moreover, the same holds if we multiply it by \textit{any} function of $s_{34}$, $s_{35}$, $s_{14}$ or $s_{15}$, e.g.
\begin{align}
\sum_{S_5} s_{12} s_{23} s_{45} \, \text{PT}(12345) \text{PT}(12354) \times f(s_{34}) = 0 \,.
\end{align}
We emphasize that in this example, the identity follows from the interplay between the symmetry properties of the rational prefactor 
and the fact that $f(s_{34})$ depends on a single variable only. It does not imply any functional identity for the latter. 
Nonetheless, these simple examples clearly show how the study of such relations is not only interesting on its own, but is crucial in order to find an elegant expression for a scattering amplitude. In this regard, it is interesting that in both the tree-level~\eqref{eq:M_tree2} and the one-loop case~\eqref{eq:M_1loop} such an elegant expression involves the sum over the permutations of a compact  `seed' function.

\section{Structure of infrared divergences and hard function}
\label{sec:divergences}

As was mentioned in the introduction, gravity amplitudes exhibit a remarkably simple infrared (IR) behavior. 
They are free of collinear singularities, and have soft divergences only \cite{Weinberg:1965nx}.
As a result, the leading IR divergence of an $\ell$-loop amplitude is $1/\epsilon^{\ell}$, compared to $1/\epsilon^{2 \ell}$ in gauge-theories.
Moreover, the soft divergences of gravity amplitudes are given by the formula
\begin{align}
\label{eq:IRfactorization}
\mathcal{M}_5(s_{ij},\epsilon) = \mathcal{S}_5(s_{ij},\epsilon) \, \mathcal{M}^f_5(s_{ij},\epsilon)\, , 
\end{align}
where the gravitational soft function $\mathcal{S}_5(s_{ij},\epsilon)$ captures all soft singularities, 
which means that $\mathcal{M}^f_5$ is finite in four dimensions.. 
The gravitational soft function is simply obtained by exponentiating the IR divergence of the one-loop amplitude~\cite{Weinberg:1965nx, Dunbar:1995ed, Naculich:2008ew, Naculich:2011ry, White:2011yy, Akhoury:2011kq,Beneke:2012xa},
\begin{align}
\mathcal{S}_5(s_{ij},\epsilon) = \exp{\left[\frac{\sigma_5}{\epsilon}\right]}\,, \qquad \qquad \sigma_5 = \left(\frac{\kappa}{2} \right)^2 \sum_{j=1}^5 \sum_{i<j} s_{ij} \log\left(\frac{-s_{ij}}{\mu^2} \right) \, ,
\end{align}
where $\mu$ is a factorization scale. 
In this sense the soft divergences of gravity amplitudes are one-loop exact. 
Letting $\epsilon \rightarrow 0$ in the finite quantity $\mathcal{M}^f_5$ defines an IR-safe hard function, or remainder function,
\begin{align}
\mathcal{H}_5(s_{ij}) \equiv \underset{\epsilon \to 0}{\lim} \, \mathcal{M}^f_5(s_{ij},\epsilon) \,.
\end{align}
Given the above discussion, the hard function is the only truly new piece of information (relevant in four dimensions).

Let use denote by $\mathcal{M}^{(\ell)}_{5;w}$ the transcendental weight-$w$ component of the $\ell$-loop five-particle amplitude. Since the one- and two-loop amplitudes have uniform transcendental weight, $\mathcal{M}^{(\ell)}_{5;w}$ corresponds to the $\mathcal{O}(\epsilon^{w-2 \ell})$ term of the $\epsilon$-expansion of $\mathcal{M}^{(\ell)}_5$ for $\ell=1,2$.  
  The tree-level hard function then coincides with the tree-level amplitude, and the one-loop correction is simply given by the order-$\epsilon^0$ terms of the one-loop amplitude  
\begin{align}
\mathcal{H}_5^{(0)} = \mathcal{M}^{(0)}_5 \,, \qquad \qquad \mathcal{H}_5^{(1)} = \mathcal{M}_{5;2}^{(1)} \,.
\end{align}  
  
The factorization formula~\eqref{eq:IRfactorization} entirely determines the IR poles by lower-order data. For example, at two loops, we have
\begin{align}
& \mathcal{M}^{(2)}_{5;0} = 0\,, \nonumber \\
& \mathcal{M}^{(2)}_{5;1} = 0\,, \nonumber \\
& \mathcal{M}^{(2)}_{5;2} = \frac{\sigma_5^2}{2} \, \mathcal{M}^{(0)}_5 + \sigma_5 \, \mathcal{M}^{(1)}_{5;1}\,, \nonumber \\
& \mathcal{M}^{(2)}_{5;3} = \sigma_5 \, \mathcal{M}^{(1)}_{5;2}   \, ,
\end{align}
and the two-loop contribution to the IR-safe hard function $\mathcal{H}_5$ is given by
\begin{align}\label{defH2}
\mathcal{H}^{(2)}_{5} = \mathcal{M}^{(2)}_{5;4}- \sigma_5 \, \mathcal{M}^{(1)}_{5;3}.
\end{align}
Our goal is to compute $\mathcal{H}^{(2)}_{5}$.


\section{Calculation of the two-loop five-graviton amplitude}
\label{sec:calculation}

\subsection{Expected structure of the result}

Before embarking on the calculation, it is worthwhile to discuss the expected structure of the result.
The recent example of the two-loop five-particle amplitude in $\mathcal{N}=4$ super Yang-Mills theory~\cite{Abreu:2018aqd,Chicherin:2018yne} 
has shown that having a prior insight in the structure of the final answer is extremely valuable when assembling an amplitude. 

It was conjectured~\cite{Chicherin:2017dob} and subsequently shown~\cite{Abreu:2018rcw,Chicherin:2018mue,Abreu:2018aqd,Chicherin:2018old} that two-loop five-particle amplitudes are given by the class of (in general non-planar) pentagon functions described in section 2. 
Specifically, all integrals contributing to such amplitudes can be reduced, in principle, to a set of pure Feynman integrals.
Let us call the latter set $ f_i^{\text{UT}}(s_{ij},\eps)$. After integration-by-parts (IBP) reduction \cite{Chetyrkin:1981qh} to this basis,
a general two-loop five-particle amplitude will have the form
\begin{align}
\mathcal{M}^{(2)}_5 =  \sum_i R_i^{(2)}(\lambda,\tilde{\lambda},\eps) \, f_i^{\text{UT}}(s_{ij},\epsilon)\, .
\end{align}
We wish to make an educated guess about the factors $R_{i}^{(2)}$.

In maximally supersymmetric theories, scattering amplitudes are often of uniform weight.
Conjecturally, this property is related \cite{ArkaniHamed:2010gh} to their four-dimensional integrands\footnote{Note that in some situations, the four-dimensional 
integrand analysis may be insufficient to determine whether an integral is UT or not. An example are Feynman integrals with numerators built from Gram determinants that vanish identically for four-dimensional loop momenta, but that may yield non-zero results after integration. See \cite{Chicherin:2018old} for recent progress on identifying pure functions through a refined, $D$-dimensional integrand analysis.}  being written 
as a {\it dlog}-form, with constant prefactors. In particular, such integrands do not have double poles.
While certain integrands in $\mathcal{N}=4$ sYM can be shown to have this property \cite{ArkaniHamed:2012nw,Arkani-Hamed:2014via},
to the best of our knowledge the situation is inconclusive in $\mathcal{N}=8$ supergravity. 
Recent work \cite{Herrmann:2018dja,Bourjaily:2018omh} analyzes in particular certain poles at infinity. While the authors find that double and higher
poles may appear in general, their work suggests that the two-loop five-point $\mathcal{N}=8$ supergravity amplitude 
is free of double poles at infinity. 
We take this as an encouraging hint that the amplitude may be of uniform weight.
If this is the case, then the rational factors $r_i^{(2)}$ would be independent of $\eps$.

Secondly, we would like to draw inspiration from the tree-level (\ref{eq:M_tree}) and one-loop amplitudes (\ref{eq:M_1loop}).
The reader familiar with $\mathcal{N}=4$ super Yang-Mills may know that in that theory, one can sometimes deduce (conjecturally) from
lower-loop results what rational factors may appear in amplitudes in general.
Due to the dimensionality of the coupling $\kappa$, the situation is different here, in that the set of factors necessarily changes with 
the loop order. Nevertheless, based on the factors present at tree-level and one-loop, we would expect the following two classes
of factors to be relevant,
\begin{align}
\label{eq:2loopPTPT}
s_{ij} s_{kl} s_{mn} s_{op} \text{PT}(\sigma) \text{PT}(\rho) \, ,
\end{align}
and
\begin{align}
\label{eq:extraPrefactor2loop}
\frac{s_{i j}}{\epsilon_5}\frac{[12][23][34][45][51]}{\vev{1 2}\vev{2 3}\vev{3 4}\vev{4 5}\vev{5 1}}\,.
\end{align}
These correspond to the class of factors encountered at one-loop, multiplied by an additional factor of $s_{ij}$.
We find that of the set (\ref{eq:2loopPTPT}), 510 are linearly independent. We choose a basis in this space, which we denote by $r_{i}^{(2)}$, with $i=1,\ldots 510$.
Further adding the factors of the type (\ref{eq:extraPrefactor2loop}) gives 5 new degrees of freedom, which can
be chosen as (\ref{eq:extraPrefactor2loop}), with $j=i+1$, and $i=1,\ldots, 5$. We denote the latter by $r_{510+i}^{(2)}$.
Note that trading two of the $s_{ij}$ in eq.~\eqref{eq:2loopPTPT} with $\epsilon_5$ does not yield additional independent objects. 

To summarize, based on the discussion of the last two paragraphs, we arrive at a refined ansatz for the form of the amplitude,
\begin{align}
\label{eq:conjecture1}
\mathcal{M}^{(2)}_5 =  \sum_{i=1}^{515} r_i^{(2)} \, f_i^{\text{UT}}(s_{ij},\epsilon)\,,
\end{align}
with the $r^{(2)}_{i}$ being independent of $\eps$.

It is useful to consider the $\eps$ expansion of the uniform weight integrals.
The two-loop integrals have in general up to fourth poles in the dimensional regulator $\eps$.
As was discussed in section 4, after summing all contributions, the two-loop amplitude is expected
to have a double pole only. We are interested in the expansion up to the finite part. This leads us to
\begin{align}
\label{eq:conjecture2}
\mathcal{M}^{(2)}_5 = \frac{1}{\epsilon^2} \sum_{j=1}^{515} r^{(2)}_j \sum_{w=0}^{2} \epsilon^w  \, g_j^{(w)} + \mathcal{O}(\eps)\, ,
\end{align}
where $g_j^{(w)}$ are weight $w$ iterated integrals (we will only need their symbols) in the pentagon alphabet of~\cite{Chicherin:2017dob}, as reviewed in section 2.

In the following, we will test the conjecture (\ref{eq:conjecture1}), (\ref{eq:conjecture2}).
This is done in two steps, first verifying the $\eps$ independence of the prefactors of the $f_i^{\text{UT}}$ in eq.~(\ref{eq:conjecture1}), and secondly computing the $r^{(2)}_j$ in eq. ~(\ref{eq:conjecture2}).



\subsection{Two-loop integrand}

\begin{figure}[t]
  \begin{center}
    \includegraphics[width=0.28\columnwidth]{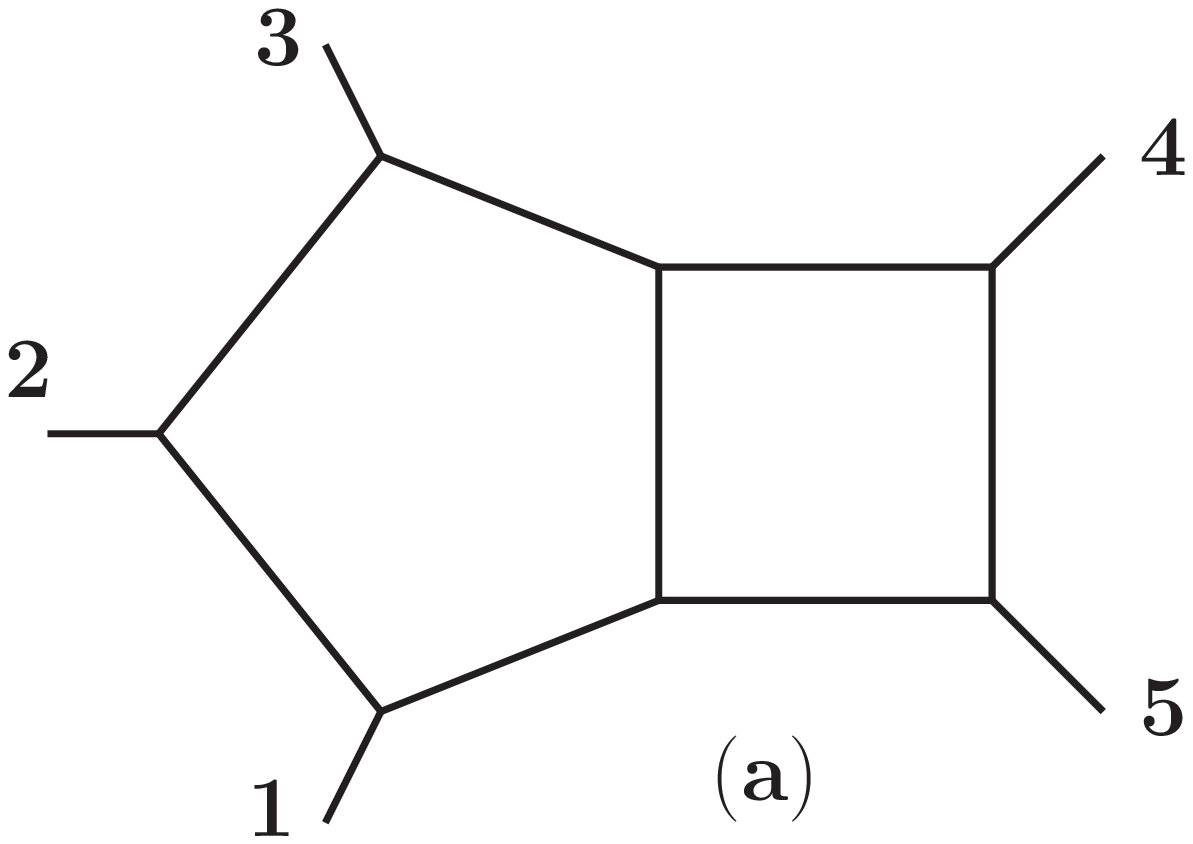}
    \includegraphics[width=0.28\columnwidth]{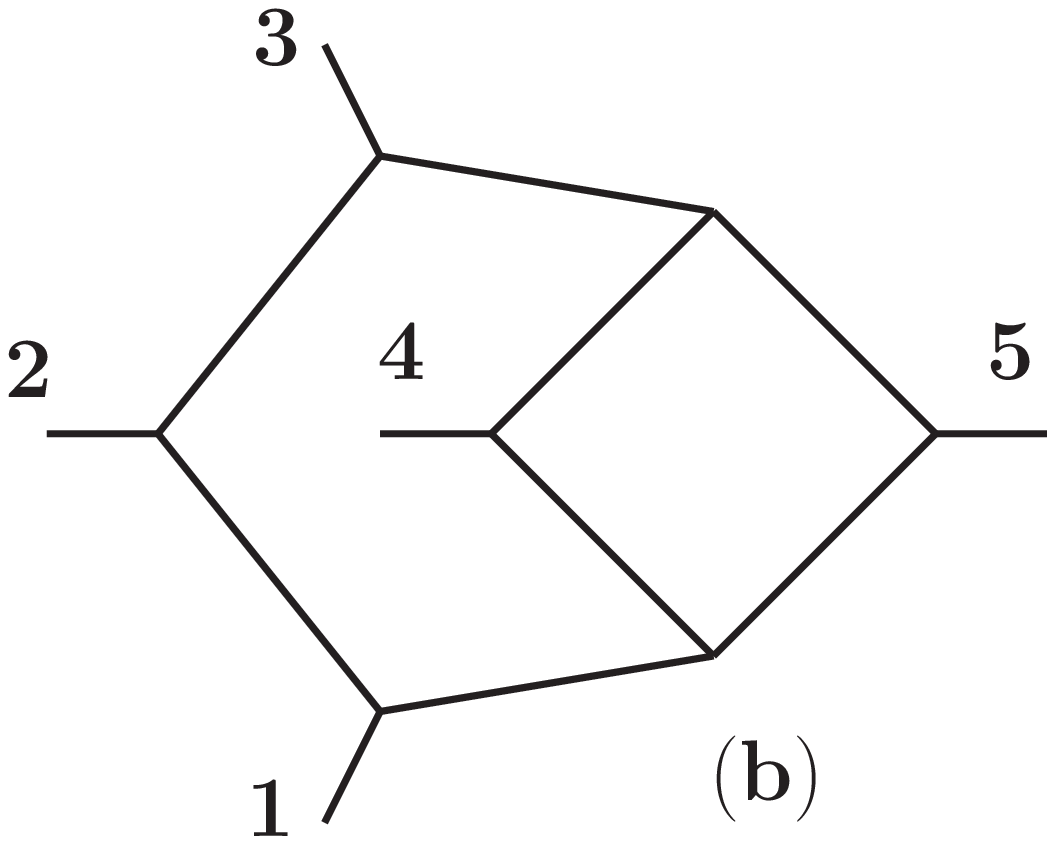}
\includegraphics[width=0.28\columnwidth]{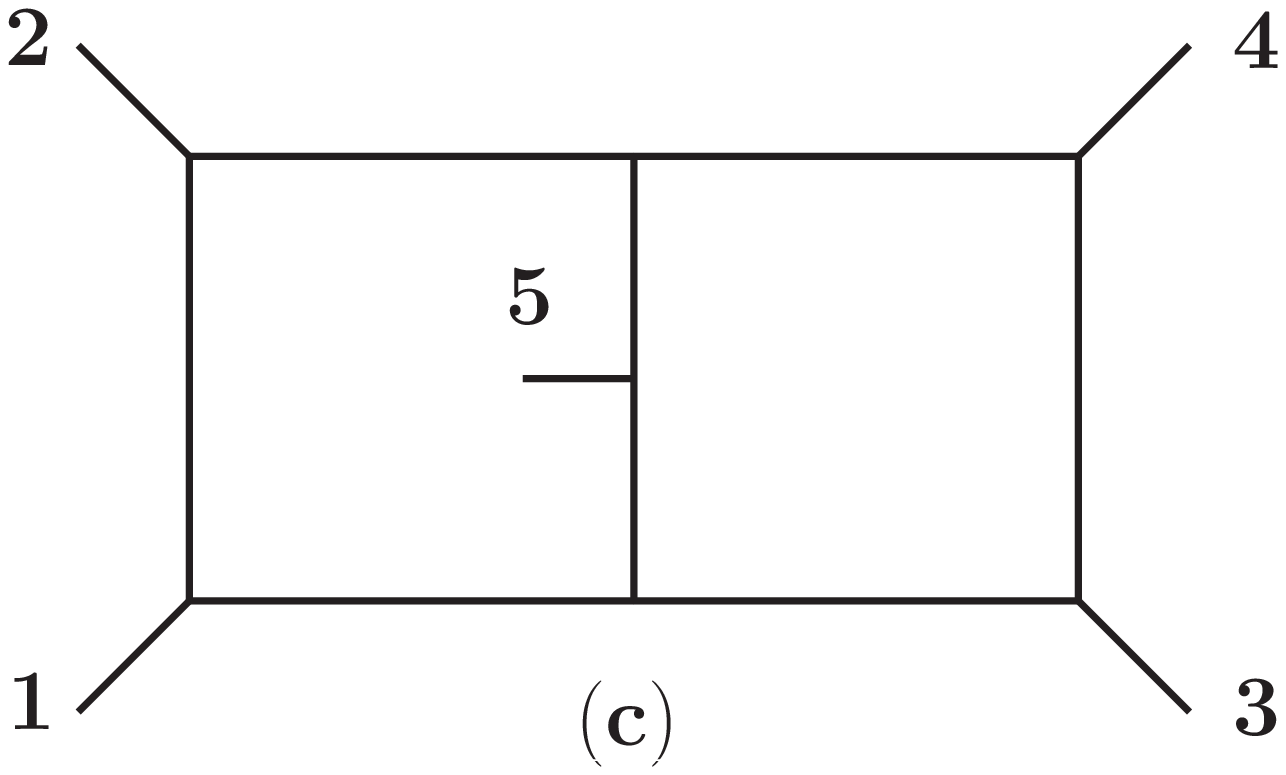}
\includegraphics[width=0.28\columnwidth]{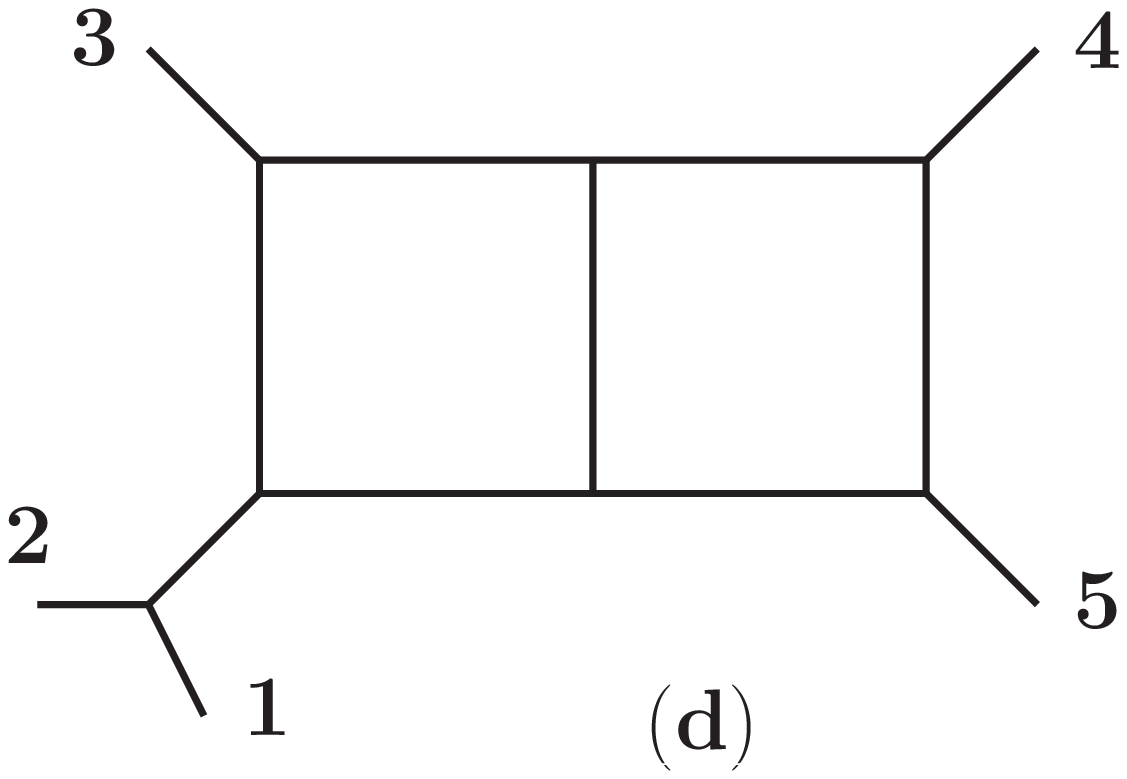}
\includegraphics[width=0.28\columnwidth]{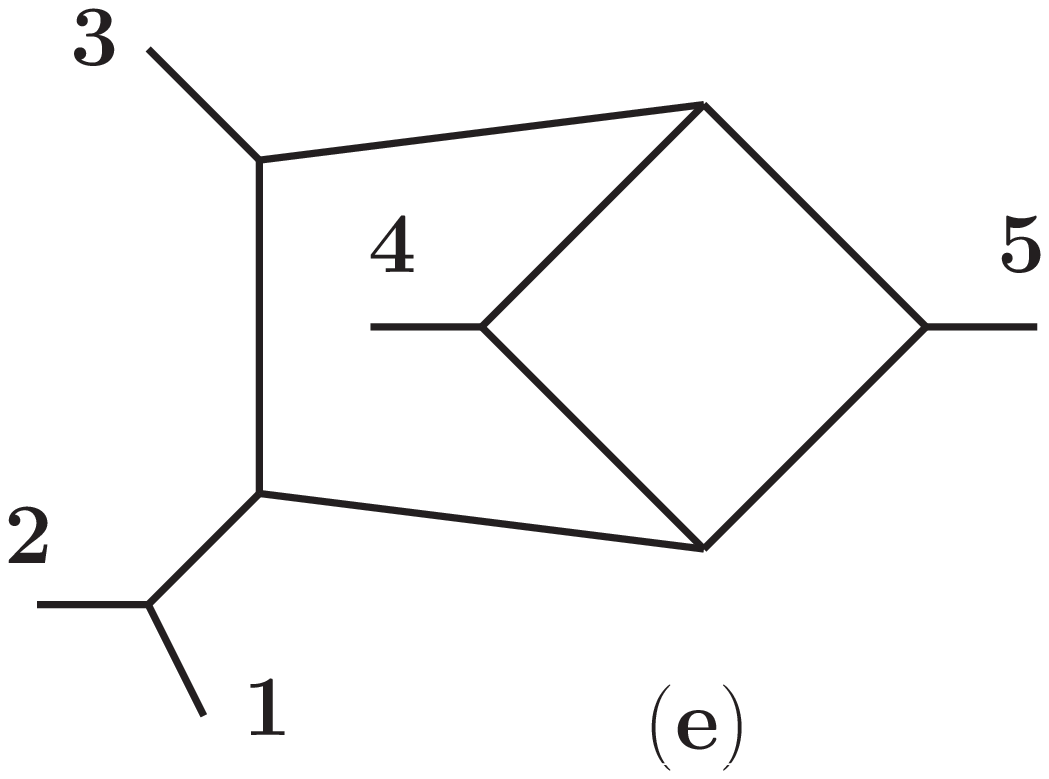}
\includegraphics[width=0.28\columnwidth]{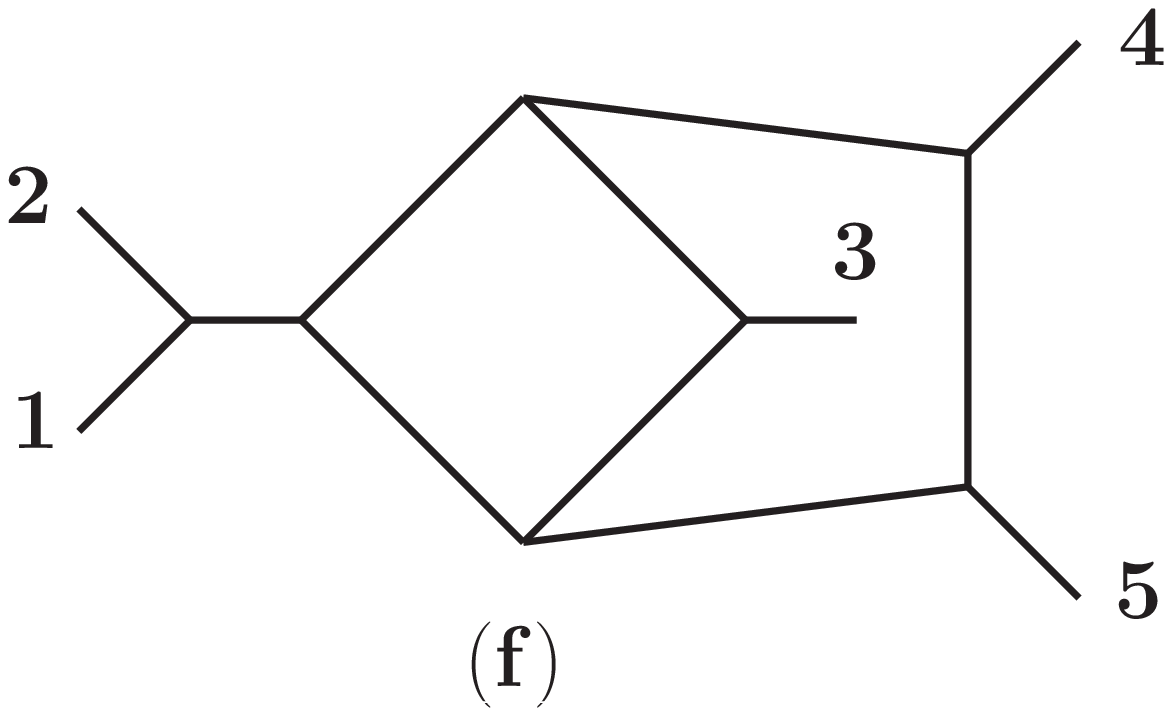}
    \caption{The six Feynmal integral topologies used to define the integrand of the two-loop five-point amplitude in
      $\mathcal{N}=8$ supergravity amplitude \cite{Carrasco:2011mn}.}
  \label{fig:sixdiagram}
  \end{center}
\end{figure}

The starting point of our calculation is the expression of the two-loop five-point $\mathcal N=8$ supergravity {\it integrand} given by ref.~\cite{Carrasco:2011mn}. 
The latter was obtained by ``double-copying" the numerators of the corresponding $\mathcal{N}=4$ super Yang-Mills integrand, in the way dictated by the color-kinematics duality~\cite{Bern:2008qj}. 
This representation is valid in the regularization scheme where the external states and momenta $p_i^{\mu}$ are four-dimensional, and the internal momenta $k_i^{\mu}$ live in $D=4-2\eps$ dimensions.
In terms of the six integral topologies shown in Fig.~\ref{fig:sixdiagram}, the supergravity amplitude is written as  
\begin{align}
\label{eq:M_CJ}
\mathcal{M}_5^{(2)} = 
 \sum_{S_5}\left(\frac{\mathcal
  I^{(a)}}{2}+\frac{\mathcal  I^{(b)}}{4}+\frac{\mathcal
  I^{(c)}}{4}+\frac{\mathcal  I^{(d)}}{2}+\frac{\mathcal  I^{(e)}}{4}+\frac{\mathcal I^{(f)}}{4} \right)\,.
\end{align}
 Here, each $\mathcal I^{(i)}$,
for $i=a,b,c,d,e,f$, is a two-loop integral
\begin{equation}
  \label{sugra_integrand}
  \mathcal I^{(i)}=\int \frac{d^D k_1}{i \pi^{\frac{D}{2}}} \frac{d^D k_2}{i \pi^{\frac{D}{2}}}
  \frac{N^{(i)}(k_1,k_2)}{D_1^{(i)} D_2^{(i)} D_3^{(i)} D_4^{(i)}
    D_5^{(i)} D_6^{(i)} D_7^{(i)} D_8^{(i)} }\,,
\end{equation}
where $D^{(i)}_j$ is an inverse propagator for the diagram $(i)$, and where $N^{(i)}$ are numerator terms,
given explicitly in ref.~\cite{Carrasco:2011mn}.

In order to calculate the amplitude, we first reduce the integrals of eq. (\ref{sugra_integrand}) to a linear combination of master integrals via integration-by-parts identities \cite{Chetyrkin:1981qh}, and further convert them to a linear combination of integrals with \textit{uniform transcendental weight}~\cite{Chicherin:2018old}.

\subsection{Uniform transcendental weight integral basis}
  \label{sec:UTbasis}

Experience shows that analytic multi-loop calculations are substantially simplified by making a good choice of integral basis, namely a basis of integrals
with \textit{uniform transcendental weight} (UT)~\cite{Henn:2013pwa}, also called pure functions. By definition, UT integrals have the very transparent analytic structure
\begin{align}
\label{eq:UTform}
\mathcal{I}^{(\ell)}_{\text{UT}}(s_{ij},\epsilon) = \frac{c}{\epsilon^{2 \ell}} \sum_{w=0}^{\infty} \epsilon^w h^{(w)}(s_{ij},\epsilon) \, ,
\end{align}
where $\ell$ is the loop-order, $c$ is a conventional normalization factor, and $h^{(w)}$ is a $w$-fold (weight-$w$) $d\log$ iterated integral \cite{Goncharov:2010jf,Duhr:2011zq}. Two-loop UT integrals have constant leading poles, transcendental weight 1 (logarithms) at order $1/\epsilon^3$, and in general weight $w$ at order $\epsilon^{w-4}$.

The UT bases of the integral families relevant for massless five-particle scattering at two loops are known: 
UT bases for the integral families $(a)$ and $(b)$ are given respectively in refs.~\cite{Gehrmann:2015bfy,Gehrmann:2018yef} 
and~\cite{Abreu:2018rcw,Chicherin:2018mue}, and the recent progress of refs.~\cite{Abreu:2018aqd, Chicherin:2018yne, Chicherin:2018old} 
means that a UT basis for family $(c)$ is available as well. 
We schematically denote the transformation between the UT basis and the master integral basis from the Laporta algorithm as
 \begin{gather}
  \label{eq:3}
  \mathbf{\tilde I}^{(a)} =T^{(a)} \cdot \mathbf{I}^{(a)}\,,\qquad
  \mathbf{\tilde I}^{(b)} =T^{(b)} \cdot \mathbf{I}^{(b)}\,, \qquad \mathbf{\tilde I}^{(c)} =T^{(c)} \cdot \mathbf{I}^{(c)}\,,\
\end{gather}
where each $\mathbf{I}^{(i)}$, $i=a,b,c$, is a vector with the Laporta master integrals of diagram $(i)$, $\mathbf{\tilde I}^{(i)}$ is a vector of the UT basis integrals, and $T^{(i)}$ is the transformation matrix. The transformation matrices $T^{(i)}$ can be easily computed by IBP reducing via the Laporta algorithm the UT basis integrals. The inverse transformation matrices $(T^{(i)})^{-1}$, which convert the Laporta master integrals to UT integrals, were computed
using the sparse linear algebra method of ref.~\cite{Boehm:2018fpv}.
  
The UT basis integrals $\mathbf{\tilde I}^{(i)}$ obey the canonical differential equations~\cite{Henn:2013pwa}
\begin{align}
\label{canonicalDEpentagon}
d\mathbf{\tilde I}^{(i)}(s_{ij};\epsilon) = \epsilon \left(
  \sum_{k=1}^{31} A_k^{(i)} d\log W_k(s_{ij}) \right)
  \mathbf{\tilde I}^{(i)}(s_{ij};\epsilon),\quad i=a,b,c \, ,
\end{align}
where $A_k^{(i)}$ is a constant rational matrix, and the $W_k$'s are symbol letters of the pentagon alphabet~\cite{Chicherin:2017dob,Chicherin:2018old}, which we reviewed in section 2.

The canonical differential equations~\eqref{canonicalDEpentagon} for the three integral families, together with the corresponding boundary values at the leading order in the $\epsilon$-expansion, are known in the literature~\cite{Gehrmann:2015bfy,Gehrmann:2018yef, Chicherin:2018mue, Abreu:2018aqd, Chicherin:2018yne, Chicherin:2018old}, and allow one to straightforwardly write down the \textit{symbol}~\cite{Goncharov:2010jf} of the UT basis integrals $\mathbf{\tilde I}^{(i)}$.

For this reason, we use the inverse transformation matrices $(T^{(i)})^{-1}$ of eq.~\eqref{eq:3} to convert the Laporta master integrals $\mathbf{I}^{(i)}$ in eq.~\eqref{sugra_integrand_IBP} to UT basis integrals $\mathbf{\tilde I}^{(i)}$. The supergravity amplitude then takes the form
\begin{gather}
  \label{sugra_integrand_UT}
  \mathcal{M}_5^{(2)} = 
  \sum_{S_5} \left(\sum_{j=1}^{61}
    \tilde c^{(a)}_j \tilde I^{(a)}_j + \sum_{j=1}^{73}
    \tilde c^{(b)}_j \tilde I^{(b)}_j+ \sum_{j=1}^{108}
    \tilde c^{(c)}_j \tilde I^{(b)}_j\right) \,,
\end{gather}
where the coefficients $\tilde c^{(i)}_j = \tilde{c}^{(i)}_j (\lambda, \tilde{\lambda},\epsilon)$ again depend on the kinematics through spinors, and on $\epsilon$. This apparent dependence on $\epsilon$ raises the question: {\it does the two-loop five-point $\mathcal N=8$
supergravity amplitude have uniform transcendental weight?} 

In the next section, we will see that after using the explicit symbol expression of the integrals, and summing over all permutations, 
the $\epsilon$ dependence coming from the coefficients $\tilde c^{(i)}_j$ cancels out, and the two-loop five-point $\mathcal N=8$
supergravity amplitude {\it does indeed} have uniform transcendental weight.

The master integrals were computed previously for $i=a$ \cite{Gehrmann:2015bfy,Papadopoulos:2015jft,Gehrmann:2018yef},
 $i=b$ \cite{Abreu:2018rcw,Chicherin:2018mue}, $i=d$ \cite{Gehrmann:2000zt} and $i=e,f$ \cite{Gehrmann:2001ck}, 
 as well as for $i=c$ at symbol level \cite{Abreu:2018aqd,Chicherin:2018old,Chicherin:2018yne}.
  
\subsection{Integration-by-parts reduction}
  \label{sec:IBPs}
In order to calculate analytically the two-loop five-point $\mathcal N=8$ supergravity amplitude, we first need to reduce the integrals in eq.~\eqref{sugra_integrand} to a linear combination of master
integrals. 

The diagrams in the first row of Fig.~\ref{fig:sixdiagram} represent the three distinct integral topologies relevant for massless
five-particle scattering at two loops. The diagrams $(d)$, $(e)$ and $(f)$ can be obtained from the top-diagrams $(a)$, $(b)$ and $(c)$, 
respectively, by pinching one internal line and supplying an extra propagator which does not depend on the loop momenta, as shown in the figure.
All these integral families have been previously calculated. There are $61$ master integrals in the family $(a)$~\cite{Gehrmann:2015bfy,Papadopoulos:2015jft,Gehrmann:2018yef}, $73$ in $(b)$~\cite{Chicherin:2018mue} (see also~\cite{Chicherin:2018ubl,Chicherin:2018wes,Abreu:2018rcw}), and $108$ in $(c)$~\cite{Chicherin:2018old,Abreu:2018aqd}.
We denote them as
\begin{align}
  \label{eq:2}
 & I^{(a)}_j, \quad j=1,\ldots, 61\,,  \\
 & I^{(b)}_j, \quad j=1,\ldots, 71\,,  \\
 & I^{(c)}_j, \quad j=1,\ldots, 108\,.
\end{align}

Note that in the representation~\eqref{sugra_integrand} of the supergravity integrand, the numerators of the diagrams $(a)$, $(b)$
and $(c)$ have degree two, namely they depend quadratically on the loop momenta. This representation
thus contains \textit{reducible} integrals in the sense of the Laporta algorithm, and therefore an IBP reduction is necessary. 

We use the IBP solvers {\sc FIRE}~\cite{Smirnov:2008iw} and {\sc Reduze2}~\cite{vonManteuffel:2012np}
to carry out the IBP reduction of~\eqref{sugra_integrand} for the particular ordering of the external legs shown in
Fig.~\ref{fig:sixdiagram}. 
We also use the private IBP code \cite{Boehm:2018fpv} to convert the master integral choices in these public programs to our convention.

As already mentioned, the integrals $\mathcal I^{(d)}$, $\mathcal
I^{(e)}$ and $\mathcal I^{(f)}$ are treated as subdiagram integrals. 
The resulting reduced amplitude has the form
\begin{gather}
  \label{sugra_integrand_IBP}
  \mathcal{M}_5^{(2)} = 
  \sum_{S_5} \left(\sum_{j=1}^{61}
    c^{(a)}_j I^{(a)}_j + \sum_{j=1}^{73}
    c^{(b)}_j I^{(b)}_j+ \sum_{j=1}^{108}
    c^{(c)}_j I^{(c)}_j\right) \,,
\end{gather}
where the coefficient functions $c^{(i)}_j = c^{(i)}_j(\lambda,\tilde{\lambda},\epsilon)$ depend on the kinematics through spinors $\lambda$,$\tilde{\lambda}$, as they have non-vanishing helicity weight, and on the dimensional regularization parameter $\epsilon$. 

Note that, since the symbols of all required master integrals are known, it is {\it not} needed to further identify master 
integral relations for other permutations of the external legs, as we can act with the permutations directly on the symbols.

\subsection{Test of expected form of the result}

We are now in a position to test the ansatz (\ref{eq:conjecture2}).
We substitute the known symbols for the UT basis integrals in eq.~\eqref{sugra_integrand_UT}, carry out the permutations while evaluating the prefactors in a random kinematic point, and sum up all terms. Note that we keep the explicit dependence in $\epsilon$. We do this for 515 different random kinematic points, single out the coefficient of each individual symbol of the amplitude, and match it against a $\mathbb{Q}$-linear combination of the 515 $r_i^{(2)}$ through finite fields methods. All prefactors are found to live in the space spanned by the assumed basis. Additional kinematic points are used to validate the result. Furthermore, we observe that the non-trivial dependence on $\epsilon$ in the prefactors $\tilde{c}^{(i)}_j$ of~\eqref{eq:M_CJ} drops out, and the amplitude has uniform transcendental weight. 
In agreement with the expected infrared structure discussed in section 2, the $1/\epsilon^4$ and $1/\epsilon^3$ poles vanish identically, 
so that the amplitude takes the form given in eq. (\ref{eq:conjecture2}).

Interestingly, we observe that parity-odd symbols only enter the amplitude in the finite part. 
This was expected, since the one-loop amplitude only depends on parity-even symbols up to its finite part, 
and --~as discussed in section 4~-- the IR divergent parts of the two-loop amplitude are completely determined 
by the tree- and one-loop amplitudes.

Next, we use eq. (\ref{defH2}) to evaluate the two-loop hard function $\mathcal{H}^{(2)}_{5}$. 
We find that the latter is again expressed as
\begin{align}
\label{eq:hardrep1}
\mathcal{H}^{(2)}_5 =  \sum_{j=1}^{510} r^{(2)}_j  \, \tilde{g}_j^{(4)} + \mathcal{O}(\eps)\, ,
\end{align}
where $\tilde{g}_j^{(4)}$ are certain weight four functions. The reconstruction of the rational factors $r^{(2)}_j$ is performed using the finite-field lifting method~\cite{vonManteuffel:2014ixa, Peraro:2016wsq}.

Two interesting simplifications take place when going from the finite part of the amplitude to the hard function. First of all, we note that the symbol of $\mathcal{H}^{(2)}_{5}$ depends only on 30 of the 31 letters of the pentagon alphabet~\cite{Chicherin:2017dob}. The letter $W_{31}$, which is present in the symbol of the amplitude $\mathcal{M}^{(2)}_{5}$, drops out from the hard function. The same property was observed for the five-particle two-loop amplitude in $\mathcal{N}=4$ super Yang-Mills theory~\cite{Chicherin:2018yne,Abreu:2018aqd}. 

Secondly, as indicated in the range of the sum of eq. (\ref{eq:hardrep1}), the rational prefactors of the form~\eqref{eq:extraPrefactor2loop} drop out from the hard function.
In other words, they enter the amplitude $\mathcal{M}^{(2)}_5$ only through the $\mathcal{O}(\epsilon)$ term of the one-loop amplitude~\eqref{eq:M_1loop}. 
As a result, all rational prefactors of the five-particle two-loop $\mathcal{N}=8$ supergravity amplitude have the simple form given by eq.~\eqref{eq:2loopPTPT}. 
As a further refinement, we observe that only a subset of these factors come with non-zero coefficients. This motivates us to look for another representation of the final answer, which we present in the next section.

\subsection{Main result for the remainder function at two loops}
\label{sec:mainresult}

In the previous section, we noticed that only a subset of the 510 rational factors of the form (\ref{eq:2loopPTPT}) contributed to the answer.
We observe that the following term
\begin{align}
r_{\rm seed} = s_{12} s_{23} s_{34} s_{45} \text{PT}(12345) \text{PT}(21435) \,,
\end{align}
has the remarkable property that, under permutations, it produces only terms that lie in the needed subset.
With this motivation in mind, we found a very compact representation of the hard function, 
\begin{align}\label{hardseed}
\mathcal{H}^{(2)}_{5} =  \sum_{S_5}   r_{\rm seed}  \, h_{5}^{(2)}  \,,
\end{align}
where $h_{5}^{(2)}$ is a pure weight four function, with both even and odd components.
This formula is our main result.
A number of comments are in order.
\begin{itemize}
\item After evaluating the sum over permutations $\mathcal{S}_{5}$, one may use identities between rational factors to reduce the total number of terms to 40. 
We provide our choice of basis for this space in an ancillary file. Remarkably, all coefficients generated by eq.~(\ref{hardseed}) have the property, just like the tree-level and one-loop coefficients, of having at most single poles at locations $\langle i j \rangle =0$.
\item Eq.~(\ref{hardseed}) is a considerable improvement over eq.~(\ref{eq:hardrep1}), as it packages the same amount of information in a single
weight four function, thereby reducing the size needed to store the expression by two orders of magnitude.
\item As explained in section 3, an equation of this type has a large moduli space, and the definition of $h_5^{(2)}$ is far from unique.
The full space of integrable even/odd weight four pentagon functions (with first and second entry conditions applied, and independent on $W_{31}$) is 3691- and 1080-dimensional, respectively\footnote{The integrable symbols were constructed by means of the Mathematica package \sc{SymBuild}~\cite{Mitev:2018kie}.}. Within this space we can construct 2402 even and 719 odd symbols $\Delta h$ which satisfy $ \sum_{S_5} r_{\rm seed} \,\Delta h =0$, and could therefore be used to modify $h_{5}^{(2)} $ in eq.~(\ref{hardseed}).
\end{itemize}
We provide the seed symbol $h_{5}^{(2)}$, separated into parity even and odd parts, as an ancillary file to this paper. For convenience of the reader, we also provide a Mathematica script that performs the sum over $\mathcal{S}_{5}$.


\section{Limits of the amplitude}
\label{sec:limits}

The exponentiation of soft divergences constituted a very strong check, especially given the fact that it involved, at the $\mathcal{O}(1/\eps)$ level, genuine five-particle functions. Futhermore, the simple UT form of the answer, and relatively few rational structures needed, make us confident in the correctness of the above answer. In order to further validate our result, we verify in this section the expected factorization properties as a graviton becomes soft, or as two gravitons become collinear.
In these limits, the five-point amplitude factorizes into the four-point amplitude times a universal function, the soft factor or the collinear splitting amplitude accordingly \cite{Bern:1998sv}. The latter do not receive quantum corrections \cite{Weinberg:1965nx,Berends:1988zp}, which means that loop corrections of (super)gravity amplitudes have much simpler soft/collinear asymptotics compared to their (super) Yang-Mills counterparts. The four-point supergravity amplitude appearing in the limit is known up to two loops from \cite{Brandhuber:2008tf,Naculich:2008ew,BoucherVeronneau:2011qv}.

\subsection{Soft limit}

Recall that we are working with the five-point {\it super}-amplitude that comprises several component amplitudes that are all related by
supersymmetry. We can restrict our attention to one of its components without loss of generality. In the following we thus consider the scattering of five gravitons with helicity configuration $1^-,2^-,3^+,4^+,5^+$.

As the momentum of one of the particles becomes soft, say $p_5 \to 0$, the five-point amplitude factorizes according to~\cite{Bern:1998sv}
\begin{align} \label{softlim}
{\cal M}^{(\ell)}_5 (1^-,2^-,3^+,4^+,5^+) \underset{p_5 \to 0}{\longrightarrow} {\cal S}(5^+) \times {\cal M}^{(\ell)}_4 (1^-,2^-,3^+,4^+) \,.
\end{align}
The leading soft factor for the positive helicity graviton is given at all orders by its tree-level expression~\cite{Weinberg:1965nx,Berends:1988zp}
\begin{align} \label{softfact}
{\cal S}(5^+) = - \frac{1}{\vev{15}\vev{54}} \left[ \frac{\vev{12}\vev{24}[25]}{\vev{25}} + \frac{\vev{13}\vev{34}[35]}{\vev{35}} \right] .
\end{align}
Note that here we consider the leading soft behavior only, and omit subleading soft operators. The latter are realized as differential operators in the spinor variables and do receive quantum corrections~\cite{Bern:2014oka}.

We find it convenient to use momentum twistor variables $Z_{i}$ to introduce a parametrization of the kinematics. 
In particular, we use the Poincar\'{e} dual of the standard momentum twistors~\cite{Hodges:2009hk}. They can be obtained from the latter by swapping the helicity spinors $\lambda \leftrightarrow \tilde\lambda$, i.e. our $Z$'s have the form  
\begin{align}
Z_i = \begin{pmatrix}
	\tilde\lambda_i^{\dot\alpha}\\
	 x_{i \, \alpha\dot\alpha} \, \tilde\lambda_i^{\dot\alpha} \\
\end{pmatrix}\,, \qquad \qquad x_{i}-x_{i+1} = \lambda_i \tilde{\lambda}_i = p_i \,.
\end{align} 
The soft limit in momentum twistor space then takes the form~\cite{Bianchi:2014gla}
\begin{align} \label{softZ}
Z_5 \to Z_4 + a_1 Z_1 + \delta \left( a_2 Z_2 + a_3 Z_3 \right) \, ,
\end{align}
where $\delta$ approaches $0$ in the limit, and the parameters $a_1,a_2,a_3$ are fixed.

In the parametrization~\eqref{softZ}, $\lambda_5 \sim \mathcal{O}(\delta)$ and $\tilde\lambda_5 \sim \mathcal{O}(1)$ as $\delta \to 0$, so that the soft factor~\eqref{softfact} diverges as ${\cal S}(5^+) \sim 1/\delta^3$ in the $\delta \to 0$ limit. The Mandelstam invariants take the form
\begin{align}
& s_{12} = \frac{s}{1+ \delta \left[ \frac{y_1}{x} + \left(1+\frac{1}{x}\right)\frac{y_1}{y_3} \right]}\,, \notag\\
& s_{23} = t \equiv s \, x \,, \notag \\
& s_{34} = \frac{s}{1+\delta\left(1+\frac{1}{x}\right)y_2(1+y_3)} \,, \notag \\
& s_{45} = \frac{y_1 s \, \delta}{1+\delta \left[ \frac{y_1}{x} + \left(1+\frac{1}{x}\right)\frac{y_1}{y_3} \right]} \,, \notag\\
& s_{15} = \frac{y_2 \, (s+t)\, \delta}{1+\delta \, y_2 \left(1+\frac{1}{x}\right) (1+y_3)}\,,
\label{sparamsoft}
\end{align}  
where $y_1,y_2$ and $y_3$ are fixed parameters which specify how the soft limit $p_5 \to 0$ is approached. Letting $\delta = 0$, i.e. $p_5 = 0$, the five-point Mandelstam invariants reduce to  the usual $s,t$ variables describing four-point scattering
\begin{align}
s_{12} \to s \,,\;\; s_{23} \to t \,,\;\; s_{34} \to s \,, \;\; s_{45} \to 0 \,,\;\; s_{15} \to 0   \,.
\end{align}

Substituting the parametrization~\eqref{sparamsoft} into the 31 letters of the pentagon alphabet $\{W_i\}_{i=1}^{31}$, and expanding them up to the leading order in $\delta$, yields a reduced 15-letter alphabet. In particular, the soft limit of the pentagon alphabet contains the sub-alphabet $\{ x,1+x,s \}$ which describes the four-point amplitude ${\cal M}_4$. Since only they can appear in the right-hand side of eq.~\eqref{softlim},  the remaining 12 letters --~${\delta}$, and those involving the non-universal parameters $y_1,y_2,y_3$~-- have to drop out after taking the soft limit. This is already a very strong check of our result. On top of that, considering the soft asymptotics of the symbol expression for the two-loop five-point amplitude, we match terms of order $1/\delta^3$ on both sides of eq.~\eqref{softlim}, and find agreement.

\subsection{Collinear limit}

We consider the collinear limit of particles $4$ and $5$, i.e. we let $p_4 = z P$ and $p_5 = (1-z) P$, with $P = p_4+p_5$. In this limit the five-point amplitude factorizes 
\begin{align}
{\cal M}^{(\ell)}_5 (1^-,2^-,3^+,4^+,5^+) \, \overset{4||5}{\longrightarrow} \, {\rm Split}^{(0)}_{-}(z;4^+,5^+) \times {\cal M}^{(\ell)}_4 (1^-,2^-,3^+,P^+) \label{colllim}
\end{align}
into a universal tree-level splitting amplitude 
\begin{align}
{\rm Split}^{(0)}_{-}(z;4^+,5^+) = - \frac{1}{z(1-z)} \frac{[45]}{\vev{45}}\,,
\end{align}
and a four-point amplitude with external momenta $p_1, p_2, p_3$ and $P$~\cite{Bern:1998sv,Bern:1998xc}. The five-particle scattering Mandelstam invariants then reduce 
to the Mandelstam invariants $s,t$ of the four-point amplitude ${\cal M}^{(\ell)}_4 (1^-,2^-,3^+,P^+)$
\begin{align}
s_{12} \to s \,,\;\; s_{23} \to t \,,\;\; s_{34} \to z\, s \,, \;\; s_{45} \to 0 \,,\;\; s_{15} \to (1-z)\,t \, .
\end{align}

Once again we resort to momentum-twistor variables to find a good parametrization of the kinematics in this limit. Similarly to the soft asymptotics~\eqref{softZ}, the collinear limit in momentum-twistor space has the form~\cite{CaronHuot:2011ky,Bullimore:2011kg}  
\begin{align}\label{collZ}
Z_5 \to Z_4  + \delta \left( a_1 Z_1 + a_3 Z_3 \right) + \delta^2 a_2 Z_2\,,
\end{align}
where $\delta \to 0$ controls the limit, and the parameters $a_1,a_2,a_3$ are fixed. In this parametrization, the Mandelstam invariants take the form
\begin{align}
&s_{12} = \frac{s}{1+\delta\left(1+\frac{1}{x}\right)\frac{1}{y}+\delta^2 \left(1+\frac{1}{x}\right)} \, , \notag\\
&s_{23} = t \equiv  s \, x \, ,\notag \\ 
&s_{34} = \frac{s z}{1+\delta \, y (1+x)(1-z)}\, , \notag \\
& s_{45} = \frac{(s+t)\delta^2}{1 + \delta\left(1+\frac{1}{x}\right)\frac{1}{y} + \delta^2 \left(1+\frac{1}{x}\right)} \, , \notag \\
& s_{15} = \frac{t (1-z)}{1+\delta \, y (1+x)(1-z)}\, , \label{sparamscoll}
\end{align}  
where $y$ is a fixed parameter which specifies how the collinear limit is approached. 

By substituting the parametrization~\eqref{sparamscoll} into the pentagon alphabet, and keeping up to the leading order in $\delta$, we find a 14-letter alphabet. Note that, as the right-hand side of eq.~\eqref{colllim} contains only the letters $\{s, x,1+x\}$, the vast majority of this 14-letter alphabet has to cancel out in the collinear limit, thus making this test very stringent. Our expression for the symbol of the two-loop five-point supergravity amplitude successfully passes this test as well, and perfectly agrees with the expected collinear behavior~\eqref{colllim}.


\subsection{Multi-Regge limit}

We now investigate the multi-Regge limit~\cite{Kuraev:1976ge,DelDuca:1995hf} of the hard function $\mathcal{H}_5$ in the physical $s_{12}$-channel
\begin{align}
s_{12} \gg s_{34} > s_{45} > 0 \, ,  \qquad  s_{23} < s_{15} < 0.
\end{align}
We control the limit through the parameter $x \to 0$, and parametrize the kinematics as
\begin{align}
s_{12} = s/x^2 \,,\;\;
s_{34} = s_1/x \,,\;\; 
s_{45} = s_2/x \,,\;\;
s_{23} = t_1 \,,\;\; 
s_{15} = t_2 \,.
\end{align}

As we already observed in~\cite{Chicherin:2018yne}, when expanded at leading order in $x$,
the pentagon alphabet becomes very simple. It reduces to 12 letters only, and factorizes into four independent alphabets: 
\begin{align}
 & \{ x \}\,, \label{eq:alph1} \\
 & \{ \kappa \}\,,\label{eq:alph2} \\
 & \{ s_1 , s_2, s_1 - s_2 , s_1 + s_2 \}\,, \label{eq:alph3} \\
 & \{ z_1, z_2 , 1 - z_1 , 1- z_2 , z_1 - z_2, 1-z_1 - z_2 \}\,, \label{eq:alph4}
\end{align}
where $\kappa$, $z_1$ and $z_2$ are defined by
\begin{align}
\kappa = \frac{s_1 s_2 }{s} \,, \qquad  t_1 = - \kappa z_1 z_2 \,, \qquad t_2 = -\kappa (1-z_1)(1-z_2) \,.
\end{align}
The pentagon alphabet therefore implies a very simple functional structure in the multi-Regge limit. The one-letter alphabets~\eqref{eq:alph1} and~\eqref{eq:alph2} simply correspond to powers of logarithms. In eq.~\eqref{eq:alph3} we recognize the alphabet of the harmonic polylogarithms~\cite{Remiddi:1999ew}, while the alphabet~\eqref{eq:alph4} encodes the two-dimensional harmonic polylogarithms~\cite{Gehrmann:2001jv}.

In order to cancel the helicity weight of the hard function, we normalize it by a squared Parke-Taylor factor, e.g. by
$\left[\text{PT}(12345)\right]^2$. We define the helicity-free hard function
\begin{align}
\widetilde{\cal H}_5 = \frac{\mathcal{H}_5}{\left[\text{PT}(12345)\right]^2}\,.
\end{align}
Then, by analyzing the symbol of the hard function $\widetilde{\cal H}_5$, we 
can reconstruct analytically its leading logarithmic asymptotics as $x \to 0$
\begin{align}
\label{eq:Hregge}
&{\widetilde{\cal H}}^{(0)}_5 \to z_1 (z_1-z_2)(1-z_2) \kappa^2 +{\cal O}(x) \, , \notag\\
&{\widetilde{\cal H}}^{(1)}_5 \to - 2 z_1 (1-z_2) \left\lbrack 3 z_1(1-z_1)- z_2(1-z_2) + 4 z_1 z_2(z_1 - z_2) \right\rbrack \kappa^3 \, \log^2 x + {\cal O}(\log x) \,, \notag\\
&{\widetilde{\cal H}}^{(2)}_5 \to \frac{1}{3} z_1 (1-z_2)\biggl[ 5 (1-z_1)z_2 \left(z_1^2+(1-z_2)^2\right) - 80 z_1 z_2 (1-z_1)(1-z_2)(z_1-z_2)  \notag\\
& \hspace{4em} 
- 43 z_1 (1-z_2)\left((1-z_1)^2 + z_2^2\right)\biggr] \kappa^4 \, \log^4 x + {\cal O}(\log^3 x) \,.
\end{align}
It is worth emphasizing that the analytic expression of leading logarithmic contributions to the hard function in the multi-Regge limit~\eqref{eq:Hregge} can be obtained from a symbol level analysis. We caution the reader that the above formulas may miss certain `beyond the symbol terms'. In principle, constants such as $\pi^2$ 
could be present, and have a different leading behavior as $x\to 0$ from the terms given above.

We provide the weight-4 symbol of the leading $x$-power-term of $\widetilde{\cal H}_5$ in an ancillary file.


\section{Conclusion and discussion}
\label{sec:conclusion}

In this paper, we computed for the first time the symbol of the two-loop five-particle scattering amplitude in $\mathcal{N}=8$ supergravity.
The calculation was more involved compared to the corresponding amplitude in $\mathcal{N}=4$ super Yang-Mills \cite{Abreu:2018aqd,Chicherin:2018yne},
which was completed very recently, since there was much less information about the 
expected structure of the answer available in the literature.

We validated the amplitude in several ways. In evaluating the two-loop amplitude, 
we verified the expected exponentiation of soft divergences. This is a particularly strong check, as it involves, at the level of the $\mathcal{O}(1/\eps)$ pole,
functions that depend genuinely on the five-particle kinematics. Moreover, we verified that our result has the correct factorisation properties as one graviton becomes soft,
or when two gravitons become collinear.

We defined an IR-subtracted amplitude, and defined a finite remainder, or hard function. 
The symbol of the latter constitutes the main result of this work.
We found that is has several remarkable properties.
We found that the hard function is given by a uniform weight-four symbol, 
and a small set of rational factors. The latter are a natural generalization of rational
factors appearing at tree-level and one-loop level.
The symbol alphabet is that of pentagon functions \cite{Chicherin:2017dob}.
Interestingly, the hard function depends on one letter less compared to the amplitude,
and likewise it contains fewer rational factors.

Moreover, we found a considerably more compact representation of the answer, cf. eq. (\ref{hardseed}).
This formula expresses the hard function in terms of a single seed function, summed over the $\mathcal{S}_{5}$ permutation symmetry.
The seed function consists of one rational factor multiplying a weight-four symbol.
The explicit answer is available in ancillary files.
We expect that this simple formula can be the starting point for many future
investigations. 

The latter compact form of the answer was obtained by analyzing identities between the 
rational factors appearing in the amplitude (in particular when considering the sum over 
permutations $\mathcal{S}_{5}$). The latter identities can be used to write several equivalent representations of the amplitude.
There are two types of identities that we find very intriguing.
Firstly, identities at the level of rational functions are reminiscent of relations that arise naturally 
from considering BCFW recursion relations \cite{Britto:2005fq,Bedford:2005yy}.
Secondly, identities involving both rational factors and symbols (iterated integrals) 
are reminiscent of the functional identities observed in the context of cluster algebras~\cite{Golden:2013xva}. 

We expanded our result in the multi-Regge limit. The explicit answer for the symbol can be found in ancillary files to this paper.
We wrote out explicitly the leading logarithmic contributions at one and two loops.
It would be interesting to explain these terms from Regge theory, along the lines of~\cite{Bartels:2012ra},
where similar terms were predicted for the four-graviton amplitude.

Another direction worth pursuing is the investigation of subleading soft theorems at loop level \cite{Cachazo:2014fwa,Bern:2014oka,Bern:2014vva}.
Our two-loop result can be used to test conjectures about the one-loop exactness of certain terms in the soft limit,
and provide invaluable data for future investigations.

\section*{Acknowledgments}
We thank V. Mitev for collaboration in early stages of the project. J.~M.~H. thanks B.~Mistlberger for helpful discussions. Y.~Z. thanks J. Boehm for help with the implementation of finite field and sparse linear algebra computations. We thank 
the HPC groups at JGU Mainz and MPCDF for support. 
This research received funding from the European Research Council (ERC) under the European Union's
Horizon 2020 research and innovation programme (grant agreement No 725110), {\it Novel structures in scattering amplitudes}.

\appendix

\bibliography{5point_refs3.bib}
\end{document}